\documentclass{elsart}

\usepackage{graphics,natbib,graphicx,psfig,amssymb}
\journal{New Astronomy}
\input psfig.sty

\begin{document}

\begin{frontmatter}

\title{A photometric study of V1193 Ori: detection of orbital and negative superhump periods}

\author[turkey]{T. Ak\corauthref{cor}},
\corauth[cor]{corresponding author.}
\ead{tanselak@istanbul.edu.tr}
\author[america]{A. Retter}
\author[australia]{and A. Liu}
\address[turkey]{Istanbul University, Faculty of Sciences, Department 
of Astronomy and Space Sciences, 34119 University, Istanbul, Turkey}
\address[america]{Pennsylvania State University, Department of Astronomy and 
Astrophysics, 525 Davey Lab., University Park, PA 16802-6305, 
USA (retter@astro.psu.edu)}
\address[australia]{Norcape Observatory, PO Box 300, Exmouth, 6707, Australia 
(asliu@onaustralia.com.au)}

\begin{abstract}
We present the results obtained from unfiltered photometric CCD observations 
of V1193 Ori made during 24 nights between November 2002 and January 2003. 
We found periods of $0.1430^{d}$, $0.1362^{d}$ and possibly $2.98^{d}$ in the data. 
The $0.1430^{d}$ period is consistent with the $1d^{-1}$ alias of the proposed 
orbital period of $P_{orb}$=$0.165^{d}$ \citep{Ringwaldetal1994,Papadakietal2004}. 
Thus and using the known relation between the orbital and  superhump periods, 
we interpret these periods as the orbital period of $0.1430^{d}$, the negative 
superhump period of $0.1362^{d}$ and the precession period of $2.98^{d}$. V1193 Ori 
can then be classified as a permanent superhump system.
\end{abstract}


\end{frontmatter}

\section{Introduction}
V1193 Orionis ($\alpha_{2000.0}$=$05^h$$16^m$$26.65^s$, 
$\delta_{2000.0}$=-$00^{\circ}$$12^{'}$$14.2^{''}$;
\citealp{Downesetal1997}), also named Hamuy's Blue Variable, was discovered 
by M. Hamuy while working on a photoelectric sequence around Seyfert 120 galaxy 
Arakelian 120 \citep{HamuyandMaza1986,Hamuyetal1986}. They estimated a mean 
V magnitude of 14.08 and mean colours of U-B=-0.82, B-V=+0.05, V-R=+0.08 and 
R-I=+0.11. \cite{HamuyandMaza1986} also reported a V magnitude variation with an 
amplitude of 0.33 mag and U-B colour variation with an amplitude of 0.23 mag. 

Preliminary spectroscopic analysis of V1193 Ori \citep{FlippenkoandEbtener1986}
revealed the presence of relatively strong $H_{\alpha}$ and $H_{\beta}$ emissions 
superposed on broad  absorption lines. In accordance with the earlier descriptions 
of the spectrum, a single spectroscopic observation made by \cite{Bondetal1987}
showed that the star has a very blue continuum and a very broad and shallow 
$H_{\beta}$ absorption with a central emission peak. \cite{Bondetal1987} also found 
irregular flickering with a peak to peak amplitude of more than 0.15 mag in the light 
curve of V1193 Ori. They suggested that the star is a member of the UX UMa class of 
cataclysmic variables (nova-like systems). Nova like systems are cataclysmic variables 
whose light curves do not have dwarf nova outbursts. This is understood by high mass 
transfer rates and thermally stable accretion discs, probably as a result of a previous 
nova eruption \cite[e.g.,][]{Warner1995,Retteretal1999}. Rapid flickering activity with 
an unusually high amplitude of 0.25 mag in the star's high speed photometry was found by 
\cite{WarnerandNather1988}. Later on, Ringwald et al. \citeyearpar{Ringwaldetal1994} 
verified the object's classification as an UX UMa star and suggested an orbital period 
of $0.165^{d}$ from the radial velocity study of the emission lines. The most recent 
CCD photometry of V1193 Ori was reported by \cite{Papadakietal2004}, who confirmed the 
orbital period of $0.165^{d}$ and the rapid flickering dominating the light curves of 
all photometric observations of the star. 

In this paper, we report on the most extensive photometric observations of V1193 Ori 
done so far, which suggest a refined orbital period, a negative superhump period and 
a precession period obtained from the periodogram analysis.

\section{Observations}
Photometric observations of V1193 Ori were made by one of us, Liu, with a 30 cm Meade 
LX200 telescope coupled to an Optec f3.3 focal reducer and an SBIG ST7E CCD camera. 
The telescope is located in Exmouth, Western Australia, and no filter was used. 
The quantum efficiency of the CCD camera is maximized near 620 nm and it is sensitive 
between 400 and 950 nm. Exposure times were 
300 sec in the first run which occurred between November 30 to December 
13, 2002 (HJD 2452609-2452622) and 180 sec in the second run which occurred between 
December 31 to January 12, 2003 (HJD 2452640-2452652). The observational log is given 
in Table 1. The observations span 24 nights (142.8 hours in total). We estimated 
differential magnitudes with respect to GSC4752-1133 (the comparison star), using 
GSC4752-1117 as the check star for which GSC magnitudes are $14.20^{m}$ and $14.10^{m}$, 
respectively. 

Differential magnitudes were calculated using aperture photometry. The mean GSC magnitude 
of the comparison star was added to the differential magnitudes to give a rough estimate 
of the visual magnitude. The light curve of V1193 Ori obtained during the observations 
is shown in Fig. 1. A part of the light curve of V1193 Ori observed on a typical night is 
presented in Fig. 2. Observational errors were estimated from the deviations of the K-C 
magnitudes from the nightly means and are typically about $0.037^{m}$ and $0.014^{m}$ 
for the first and second runs, respectively.

\begin{table}
\caption[]{Journal of the photometric observations. N denotes the number of observations.}
\begin{center}
\tiny
\begin{tabular}{lcccc}
\hline
Date          & HJD Start     & Duration & Expusure   & N \\
              & (HJD-2452600) & (hours)  &  (Seconds) &   \\
\hline
Nov. 30, 2002 & 09.0918     &  6.4     & 300     & 37   \\
Dec. 01, 2002 & 10.0925     &  5.9     & 300     & 38   \\
Dec. 02, 2002 & 11.0934     &  5.9     & 300     & 38   \\
Dec. 03, 2002 & 12.0593     &  6.5     & 300     & 62   \\
Dec. 04, 2002 & 13.0488     &  6.6     & 300     & 71   \\
Dec. 05, 2002 & 14.0601     &  6.5     & 300     & 70   \\
Dec. 06, 2002 & 15.0591     &  7.2     & 300     & 76   \\
Dec. 07, 2002 & 16.0751     &  6.4     & 300     & 70   \\
Dec. 09, 2002 & 18.1664     &  3.8     & 300     & 41   \\
Dec. 10, 2002 & 19.1175     &  5.5     & 300     & 60   \\
Dec. 11, 2002 & 20.0835     &  6.0     & 300     & 65   \\
Dec. 12, 2002 & 21.0531     &  6.8     & 300     & 69   \\
Dec. 13, 2002 & 22.0550     &  6.6     & 300     & 72   \\
Dec. 31, 2002 & 40.0393     &  6.4     & 180     & 95   \\
Jan. 01, 2003 & 41.0207     &  6.0     & 180     & 100  \\
Jan. 02, 2003 & 42.0218     &  6.6     & 180     & 110  \\
Jan. 03, 2003 & 43.0243     &  6.6     & 180     & 109  \\
Jan. 04, 2003 & 44.0121     &  6.6     & 180     & 108  \\
Jan. 05, 2003 & 45.0285     &  6.6     & 180     & 106  \\
Jan. 06, 2003 & 46.1301     &  1.6     & 180     & 22   \\
Jan. 09, 2003 & 49.1029     &  4.3     & 180     & 69   \\
Jan. 10, 2003 & 50.0294     &  6.0     & 180     & 98   \\
Jan. 11, 2003 & 51.0067     &  6.0     & 180     & 100  \\
Jan. 12, 2003 & 52.0344     &  6.0     & 180     & 100  \\
\hline
\end{tabular}
\end{center}
\end{table}

 \begin{figure}
 \vspace{5.8cm}
 \includegraphics[width=0.9\textwidth]{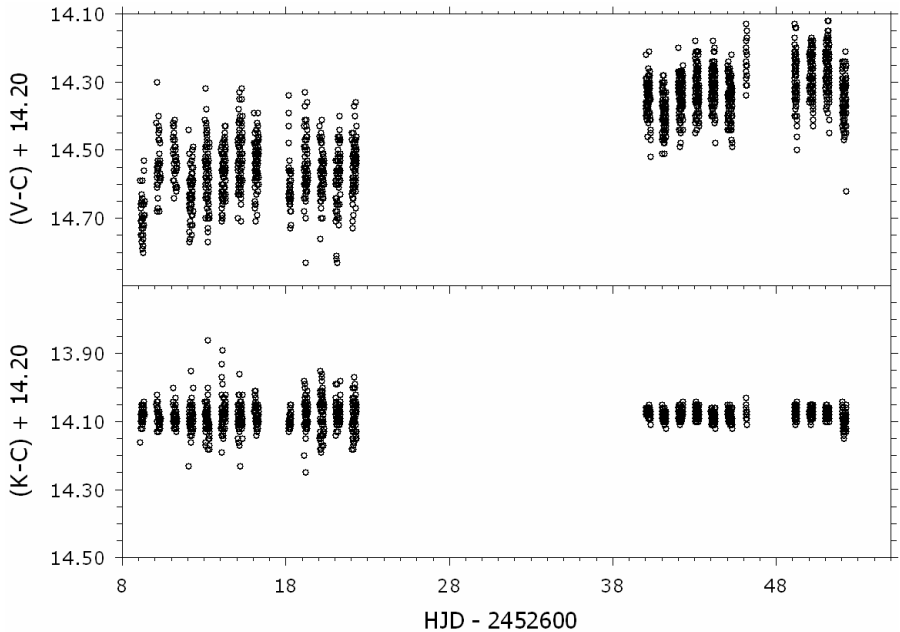}
 \caption[]{The light curve of V1193 Ori during the observing 
            run. "V" denotes unfiltered magnitude of the variable, 
            "C" and "K" unfiltered magnitudes of the comparison 
            star and the check star, respectively. The mean GSC 
            magnitude of the comparison star was added 
            to the differential magnitudes to give a rough estimate 
            of the visual magnitudes of V1193 Ori and the check star.}
 \end{figure}

\section{Analysis}
The period analysis was performed using the Data Compensated Discrete Fourier 
Transform (DCDFT, \citealp{FerrazMello1981, Foster1995}), including the CLEAN 
algoritm \citep{Roberstsetal1987}. The DCDFT method is based on a least-square 
regression on two trial functions, sin(ft) and cos(ft), and a constant. 
We note that the signal to noise ratio of the power spectra calculated by DCDFT are 
much higher than 
those of Scargle's modified periodogram \citep[see Figs.(1b) and (1c) of][]{Foster1995}.
Here f denotes the frequency. In the period analysis, we assume that the frequency, say $f_{1}$, 
that corresponds to the highest peak in the power spectrum is real and subtract its 
fit from tha data. Then, we find the highest peak, say $f_{2}$, in the power spectrum of 
the residuals, subtract $f_{1}$ and $f_{2}$ simultaneously from the raw data and 
calculate a new power spectrum etc. until the strongest residual peak is below a 
given cutoff level. To select the peaks, we also quantitavely estimated the statistical 
significances of peaks in the periodograms \citep[see,][]{Scargle1982,HorneandBaliunas1986}
and considered only those peaks of the power spectrum whose height was above the confidence level. 
In order to calculate a confidence level, we followed a conservative approach which 
is similar to the method described by \citet{Bregeretal1993} \citep[see also,][]{Kuschnigetal1997}
who gave a good criterion for the significance of a peak in the power spectrum. 
In Breger's method, the peaks in the power spectrum which are higher than the signal 
to noise ratio, S/N, of 4.0 for the amplitude are indicators of real signals. 
In order to assign a confidence level to the power spectra, we calculated the standard error 
($\sigma$) of the power values between the frequencies for which no strong 
peaks appear. We assumed 4$\sigma$ to be the confidence level for the power.
Note that we also searched for periodic brightness 
modulations by Period98 \citep{Sperl1998}, which is based on a least-square regression 
on a trial function, sin(ft), along with a zero point. 
Period98 is a program to search for and fit sinusoidal patterns within a time series 
of data in which one suspects periodic behavior. It was developed to solve problems of 
large astronomical data sets containing huge gaps. Period98 uses the discrete Fourier 
transform algorithm to calculate the frequency spectra. We found very similar power spectra 
from both techniques. We calculated the error in a frequency from the half width at the 
half maximum of the peak in the raw spectrum which is a good rough estimator of 
the uncertainty in a frequency.

 \begin{figure}
 \vspace{5.8cm}
 \includegraphics[width=0.9\textwidth]{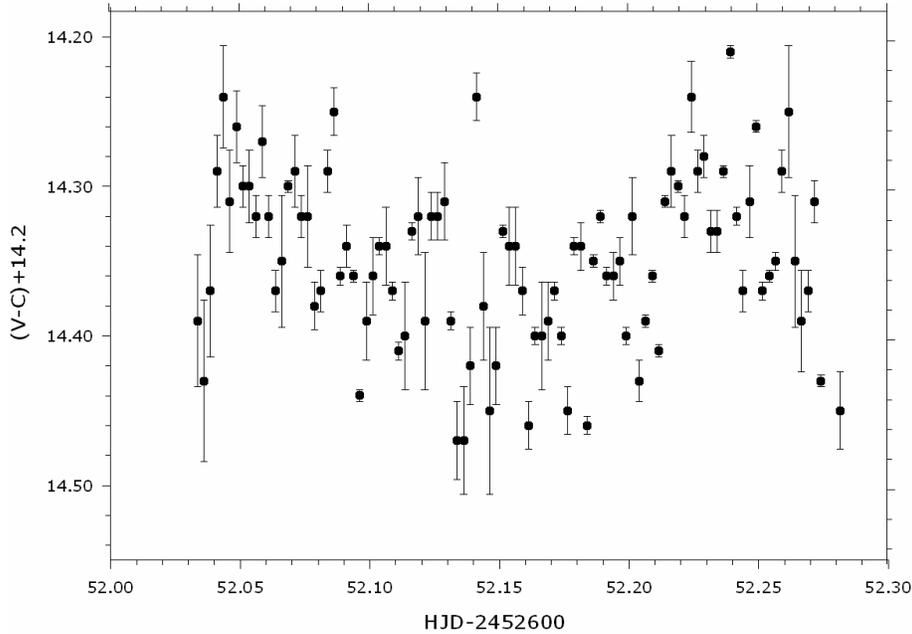}
 \caption[]{A part of the light curve of V1193 Ori observed on January 12, 2003.}
 \end{figure}

\subsection{The raw data}

The mean magnitudes of the system were estimated as 14.566$\pm$0.003 and 
14.324$\pm$0.002 mag for the first (first 13 nights, HJD 2452609-2452622)
and second (last 11 nights, HJD 2452640-2452652) runs, respectively. The error values 
represent one standard deviation of the mean values. The mean magnitude of V1193 Ori 
during the second run was 0.242 mag brighter than the first run. This magnitude 
difference between the two observing runs creates strong low frequency signals in the 
power spectrum. Therefore, we subtracted the mean magnitudes of each observing run 
from the observations of V1193 Ori in order to remove these low frequency peaks 
from the power spectra. 

The power spectrum of the light curve of V1193 Ori is dominated by the aliases of a 
low frequency signal and the observational gaps of about 1 day as demonstrated in Fig. 3 
and Fig. 4a. The strongest peak in the power spectrum is found at the frequency  
$f_{1}$=0.252$\pm$0.006 c/d ($3.97^{d}$$\pm$0.09). To search for additional signals, 
this signal and the signals originated from observational gaps were subtracted from the 
data. In the power spectrum of the residuals (Fig. 4b), the strongest peaks correspond 
to the frequencies of $f_{2}$=0.336$\pm$0.008 c/d ($2.98^{d}$$\pm$0.07), 
$f_{3}$=6.998$\pm$0.008 c/d ($0.1429^{d}$$\pm$0.0002) and $f_{4}$=7.342$\pm$0.008 c/d 
($0.1362^{d}$$\pm$0.0002) which are statistically significant.
In order to find a confidence level for the power, we calculated the standard error 
($\sigma$) of the power level to be 0.87 between 10-100 $d^{-1}$. By considering this 
standard error as the noise level, we calculated the limiting confidence level to be 
4$\sigma$=3.48 for the power, as described above.
The presence of the $2.98^{d}$ period is somewhat questionable since it is close to $3^{d}$. 
However, the data contain successive nights and there are no clear gaps of $3^{d}$. This fact 
supports the presence of this period.

 \begin{figure}
 \vspace{6.4cm}
 \includegraphics[width=0.9\textwidth]{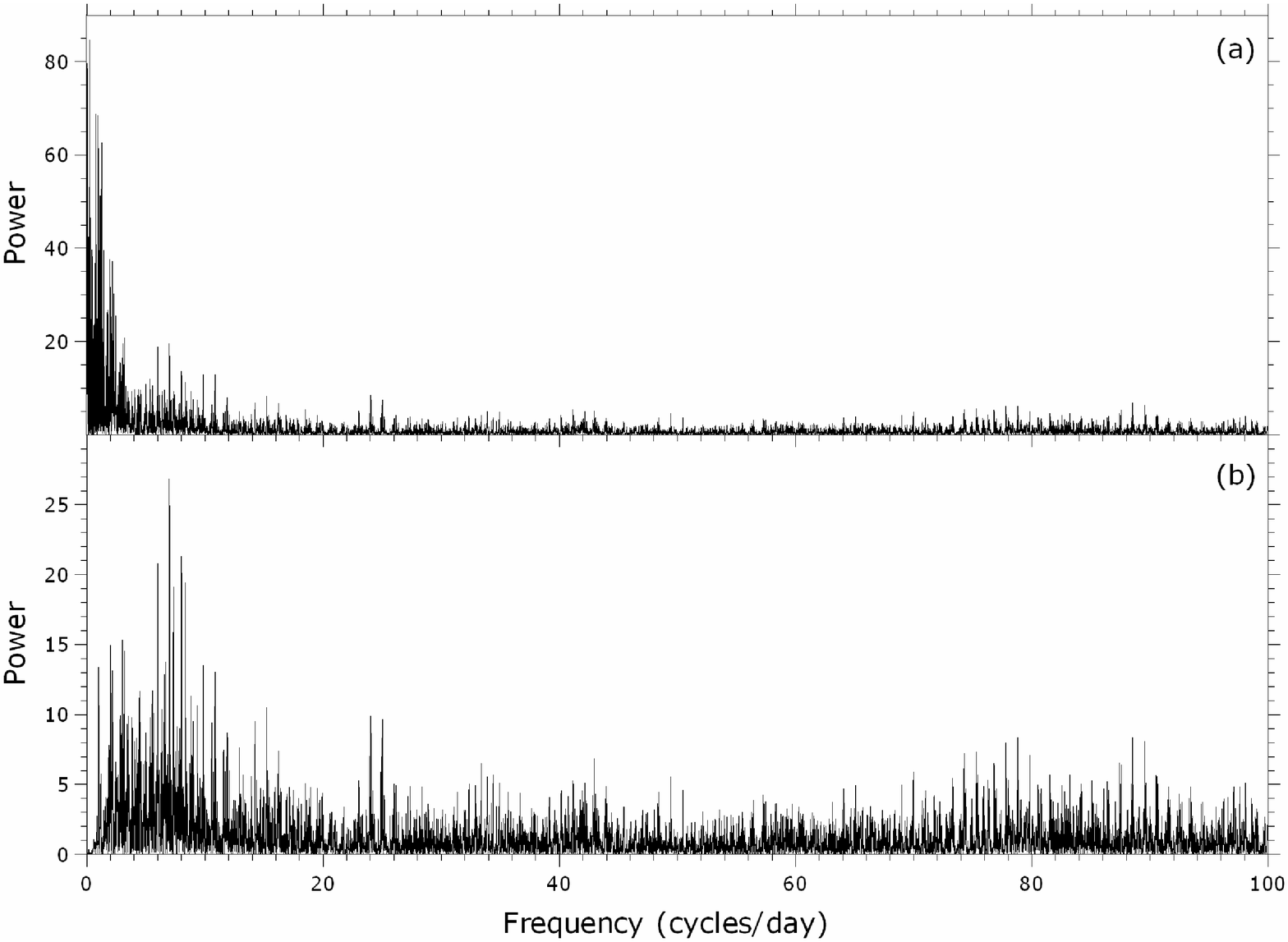}
 \caption[]{$\bf (a)$ The power spectrum of the light curve of 
                      V1193 Ori after the mean magnitudes of both 
                      observing run were subtracted from the 
                      observations. 
	      $\bf (b)$ The power spectrum of the light curve of 
                      V1193 Ori after the mean magnitudes of each 
                      night were subtracted from the observations.}
 \end{figure}

 \begin{figure}
 \vspace{6.4cm}
 \includegraphics[width=0.9\textwidth]{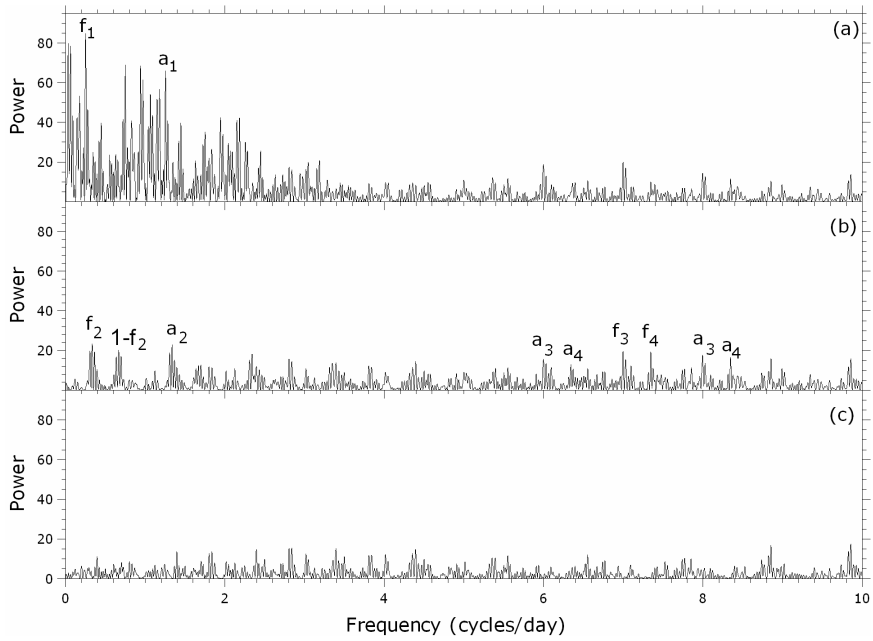}
 \caption[]{Power spectra of V1193 Ori zoomed into the frequencies 
            between 0 and 10 c/d after the mean magnitudes of 
            both observing run were subtracted from the 
            observations. '$a_{i}$' (i=1-4) represent $1d^{-1}$ 
            aliases of '$f_{i}$'. 
            $\bf (a)$ The raw power spectrum. 
            $\bf (b)$ The power spectrum after fitting and 
                       subtracting $f_{1}$=0.252 c/d. 
            $\bf (c)$ The cleanest power spectrum after fitting 
                      and subtracting $f_{1}$=0.252, $f_{2}$=0.336, 
                      $f_{3}$=6.998 and $f_{4}$=7.342 c/d.}
 \end{figure}

 \begin{figure}
 \vspace{6.0cm}
 \includegraphics[width=0.9\textwidth]{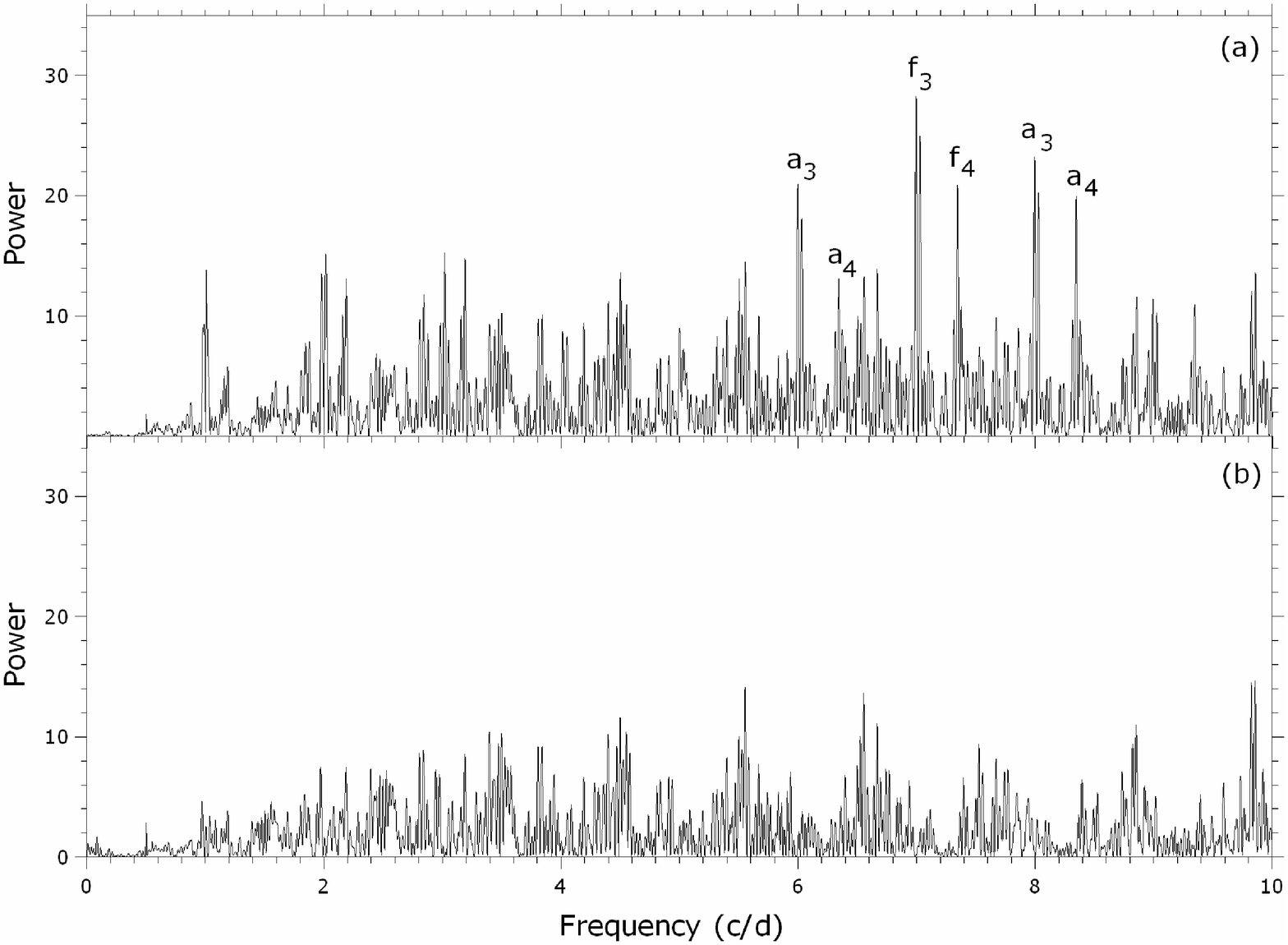}
 \caption[]{Power spectra of V1193 Ori zoomed into the frequencies 
            between 0 and 10 c/d after the mean magnitudes of 
            each night were subtracted from the observations.
            '$a_{i}$'s are as in Fig.4. 
            $\bf (a)$ The power spectrum of V1193 Ori. 
            $\bf (b)$ The cleaned power spectrum after fitting 
                      and subtracting $f_{3}$ and $f_{4}$.}
 \end{figure}

\subsection{The frequencies near 7 c/d}

To test the reliability of the peaks found near 7 c/d, we 
calculated the mean magnitudes of each night and subtracted them from 
the observations of V1193 Ori. These power spectra are shown in Fig. 3b and 
Fig. 5. The strongest peak in the power spectrum corresponds to the 
frequency 6.993$\pm$0.007 c/d ($0.1430^{d}$$\pm$0.0001) which is consistent 
with $f_{3}$ found in the de-trended data. After fitting and subtracting this 
frequency, the strongest peak in the power spectrum of the residuals is found at the 
frequency 7.342$\pm$0.008 c/d ($0.1362^{d}$$\pm$0.0001) which is consistent with 
$f_{4}$ mentioned above. 
We calculated the standard error ($\sigma$) of the power level to be 1.13 between 
10-100 $d^{-1}$. By considering this standard error as the noise level, we calculated 
the limiting confidence level to be 4$\sigma$=4.52 for the power, as described above.
The light curve folded on the $0.1430^{d}$ period is presented in Fig.7a. 
We also fitted and subtracted a sinusoid with the $0.1430^{d}$ period from the data.
The residual light curve was folded on the $0.1362^{d}$ period and is shown in Fig. 7b.

It should be noted that we also analyzed the data obtained in the two observing runs 
(see Section 2) separately and found that the periods mentioned above appear in both. 
The power spectra of the two observing runs are shown in Fig. 6.

 \begin{figure}
 \vspace{6.4cm}
 \includegraphics[width=0.9\textwidth]{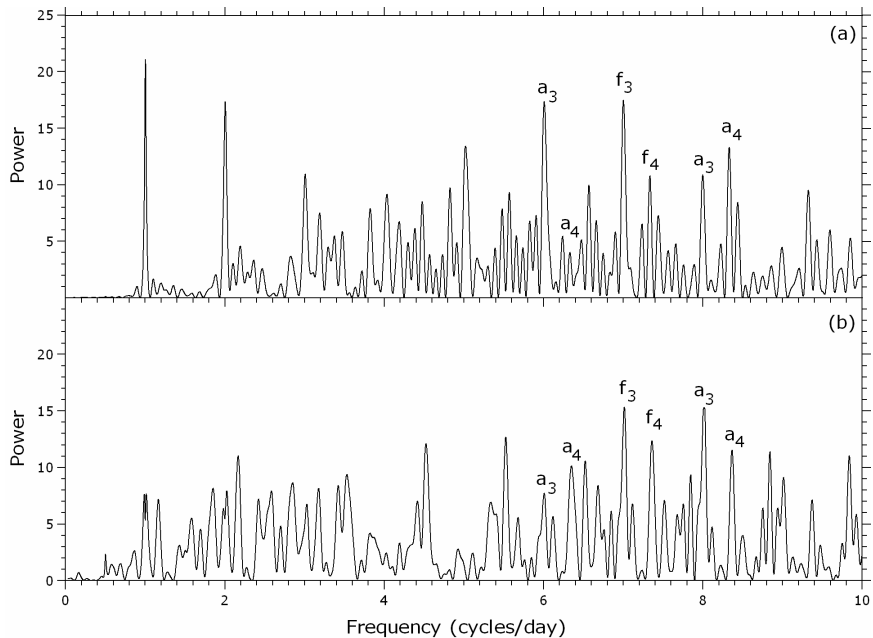}
 \caption[]{Power spectra of the two observing runs zoomed 
            into the frequencies 0-10 c/d after the mean magnitudes 
            of each night were subtracted from the observations. 
            '$a_{i}$'s are as in Fig.4. 
            $\bf (a)$ The power spectrum of the first observing 
                      run (HJD 2452609-2452622). 
            $\bf (b)$ The power spectrum of the second observing 
                      run (HJD 2452640-2452652). 
            Both graphs show the presence of $f_{3}$ and $f_{4}$.}
 \end{figure}

\section{Discussion and conclusions}

The peak that corresponds to the frequency 6.993 c/d ($0.1430^{d}$) in the power 
spectrum of V1193 Ori is consistent with the $1d^{-1}$ alias of the previously suggested 
orbital period 
($P_{orb}$=$0.165^{d}$, Ringwald et al., \citeyear{Ringwaldetal1994}; \citealp{Papadakietal2004})
Although the difference between the two corresponding frequencies is 0.932 c/d, 
not exactly 1, this difference can be ascribed to the large error in the period 
determination from the radial velocity curve which we estimate as $\pm$0.060 c/d 
by reconstructing the power spectrum of the radial velocity measurements given 
in Ringwald et al. \citeyearpar{Ringwaldetal1994}. In the power spectrum of the 
radial velocities taken from Ringwald et al. \citeyearpar{Ringwaldetal1994}, 
the $1d^{-1}$ alias of $\sim$7 c/d is about 1.5 lower 
in power than the $\sim$6 c/d peak. According to \cite{Scargle1982} the significance 
decreases exponentially, so 1.5 in power corresponds to a factor of $\sim$4.5. This 
means that the $\sim$7 c/d peak is $\sim$4.5 times lower in significance than the $\sim$6 c/d 
peak or that there is a chance of 18$\%$ (1/(4.5+1)=0.18) that the correct period 
is $\sim$7 c/d rather than $\sim$6 c/d.  In our data the $\sim$7 c/d peak is $\sim$5 
higher than the $\sim$8 c/d and $\sim$6 higher than the $\sim$6 c/d peak, so the 
chances that they represent the correct period are negligible (less than 1$\%$). 
Since most nights in our observing runs are successive and much longer than those in 
the previous studies, our results should be regarded more reliable. Thus, the period 
of $0.1430^{d}$ found from the periodogram analysis of the photometric observations of 
V1193 Ori is naturally explained as the orbital period of the system. The orbital 
period of $0.1430^{d}$ is above the period gap, which is located roughly between 2-3 h in 
the orbital period distribution of cataclysmic variables. Such an orbital period can be 
expected for V1193 Ori since most orbital periods of nova-like systems are above the 
period gap \citep{Warner1995}.

The 7.342 c/d frequency corresponds to a period of $0.1362^{d}$$\pm$0.0001, which is 
$\sim$5$\%$ shorter than the suggested orbital period of $0.1430^{d}$. Negative superhump 
periods are a few percent shorter than the orbital periods. The longer the orbital period 
is the larger the negative superhump deficit is \citep{Patterson1999}. 
The $0.1362^{d}$ period fits this relation and thus can be understood as the negative 
superhump period in V1193 Ori. Negative superhumps are explained as the beat between the 
orbital period and the nodal precession of the disc. 
The aspect at which the accretion disk is seen from Earth is modulated with the 
precession period, giving brightness modulations with $P_{pr}$ \citep{Stanishevetal2002}. 
The precession periods are typically a few days \citep{Larwoodetal1996,Woodetal2000}.
In this model, the relation between the orbital period, $P_{orb}$, the negative superhump 
period, $P^{-}_{sh}$, and the precession period of the disc, $P_{pr}$, is given by 
$1/P_{pr}$=$1/P^{-}_{sh}$$-$$1/P_{orb}$, that is $f_{pr}$=$f^{-}_{sh}$$-$$f_{orb}$. By 
choosing $f_{orb}$=6.993 c/d and $f^{-}_{sh}$=7.342 c/d from the power spectrum of V1193 Ori, 
we calculate a precession frequency of $f_{pr}$=0.35$\pm$0.02 c/d. Thus, we expect a signal 
with the frequency of $f_{pr}$=0.35 c/d in the light curve of V1193 Ori. This precession 
frequency is in good aggreement with the signal found from the de-trended data at 
$f_{2}$=0.336$\pm$0.008 c/d. So, we conclude that the precession period of the accretion 
disc in V1193 Ori is $P_{pr}$=$2.98^{d}$$\pm$0.07.

 \begin{figure}
 \vspace{6.4cm}
 \includegraphics[width=0.9\textwidth]{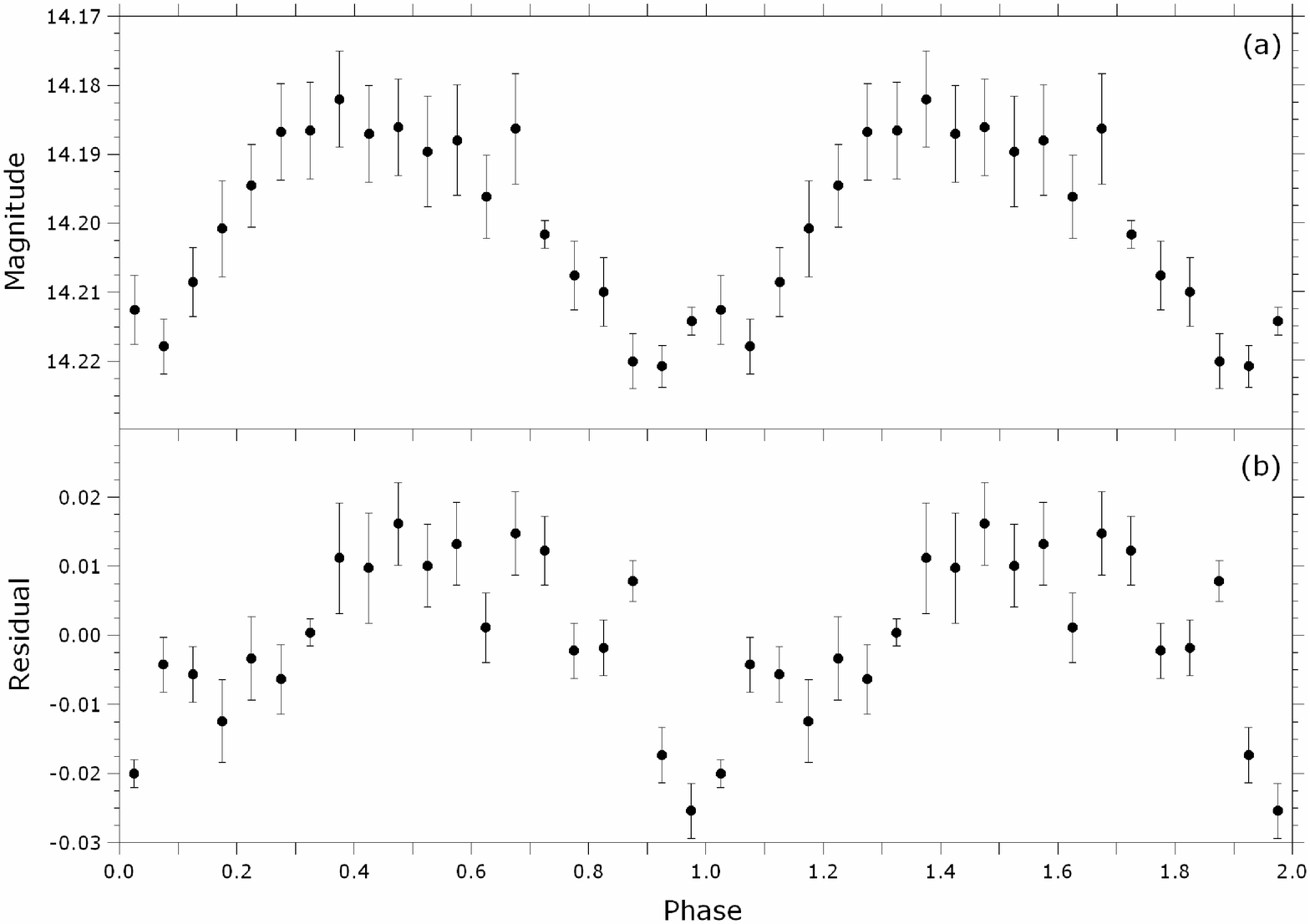}
 \caption[]{The folded light curves. The data used here are the observations 
            after the nightly means were subtracted (see Section 3.2). 
            Error bars represent one standard deviation of the mean values.
            $\bf (a)$ The light curve folded on the $0.1430^{d}$ 
                      period and binned into 20 bins. 
            $\bf (b)$ The residual light curve folded on the 
                      $0.1362^{d}$ period and binned into 20 bins 
                      after the $0.1430^{d}$ period was subtracted from 
                      the data.}
 \end{figure}

Positive superhumps are also observed in cataclysmic variables. They are explained 
as the beat between the binary motion and the precession of the disc in the apsidal 
plane \citep{Patterson1999}. The presence of permanent positive superhumps 
indicates a high and roughly constant mass transfer rate \citep[e.g.,][]{Woodetal2000}. 
Although positive superhumps are common in SU UMa-type dwarf novae, they have been found in a few 
novae and nova-like systems as well \citep{RetterandNaylor2000}, e.g., V603 Aql 
\citep{Pattersonetal1993, Pattersonetal1997}, V795 Her \citep{PattersonandSkillman1994}, 
V1974 Cyg \citep{Retteretal1997}. Positive superhumps provide good constraints 
on the mass-radius relation for the secondary stars in cataclysmic variables 
\citep{Patterson1998,Patterson2001,Pattersonetal2003}.
In a few systems positive superhumps appear simultaneously with the negative superhumps 
\citep[e.g.,][]{Pattersonetal1997,Arenasetal2000,Retteretal2003}. 
They can also be found in Algol systems \citep{Retteretal2005}.
The positive superhump periods are a few percent longer than the orbital periods. 
Patterson \citep[see also,][]{Retteretal2002} found 
that the period deficits in negative superhumps are about half the period excesses in 
positive superhumps : $\epsilon_{-}$$\approx$$-$0.5$\epsilon_{+}$ , where 
$\epsilon$=($P_{sh}$$-$$P_{orb}$)/$P_{orb}$. We found a negative superhump period 
deficit of $-0.048$ for V1193 Ori, which is appropriate for its orbital period 
\citep{Patterson1999}. From the relation between the negative superhump 
deficit and the positive superhump excess mentioned above, we expect a positive superhump 
excess of about $+0.096$ for V1193 Ori. This yields a positive superhump frequency of 
about $6.38$ c/d. From our data we put an upper limit on the amplitude of a possible positive superhump 
of about 0.037 and 0.014 mag for the first and second runs respectively.
We could not find any evidence for such a frequency in the power spectra of the 
light curve. Thus, we conclude that the system did not show a positive superhump in our 
observations or that its amplitude was below our detection limit.
However, it should be kept in mind that permanent superhumps are variable. Sometimes we 
see only a positive superhump, sometimes only a negative superhump, and other times both 
\citep[e.g.,][]{PattersonandRichman1991,Pattersonetal1993,Pattersonetal1997,Retteretal2003}.

These are the first determinations of the superhump and the possible precession in the 
light curve of V1193 Ori. Further observations are needed to confirm our results.

\section{Acknowledgements}

We thank the anonymous referee for a thorough report and useful comments that helped 
improving an early version of the paper. Part of this work was supported 
by the Research Fund of the University of Istanbul, Project Number: BYP-723/24062005.
This work was also partially supported by a postdoctoral fellowship from Penn 
State University.


\begin{thebibliography}{}

\bibitem[Arenas et al. (2000)]{Arenasetal2000}Arenas, J., Catalan, M.S., Augusteijn, T. 
$\&$ Retter, A., 2000, MNRAS 311, 135
\bibitem[Bond et al. (1987)]{Bondetal1987}Bond, H.E., Grauer, A.D., Burstein, D. $\&$ 
Marzke, R.O., 1987, PASP 99, 1097
\bibitem[Breger et al. (1993)]{Bregeretal1993}Breger, M., Stich, J., Garrido, R., Martin, B., 
Jiang, S.Y., Li, Z.P., Hube, D.P., Ostermann, W., Paparo, M., $\&$ Scheck, M. 1993, A$\&$A, 271, 482
\bibitem[Downes, Webbink $\&$ Shara (1997)]{Downesetal1997}Downes, R.A., Vebbink, R.F. $\&$ 
Shara, M.M., 1997, PASP, 109, 345
\bibitem[Ferraz-Mello (1981)]{FerrazMello1981}Ferraz-Mello, S., 1981, AJ, 86, 619
\bibitem[Filippenko $\&$ Ebtener (1986)]{FlippenkoandEbtener1986}Filippenko, A.V. $\&$ Ebtener, K., 
1986, IAUC No.4190 
\bibitem[Foster (1995)]{Foster1995}Foster, G., 1995, AJ, 109, 1889
\bibitem[Hamuy $\&$ Maza (1986)]{HamuyandMaza1986}Hamuy, M. $\&$ Maza, J., 1986, IBVS No.2867
\bibitem[Hamuy, Maza $\&$ Ruiz (1986)]{Hamuyetal1986}Hamuy, M., Maza, J., Ruiz, M.T., 1986, 
IAUC No.4172
\bibitem[Horne $\&$ Baliunas (1986)]{HorneandBaliunas1986}Horne, J.H. $\&$ Baliunas, S.L., 1986, 
ApJ 302, 757 
\bibitem[Kuschnig et al. (1997)]{Kuschnigetal1997}Kuschnig, R., Weiss, W.W., Gruber, R., Bely, P.Y. 
$\&$ Jenkner, H., 1997,  A$\&$A 328, 544 
\bibitem[Larwood et al. (1996)]{Larwoodetal1996}Larwood, J.D., Nelson, R.R., Papaloizou, J.C.B. 
$\&$ Terquem, C., 1996, MNRAS 282, 597
\bibitem[Papadaki et al. (2004)]{Papadakietal2004}Papadaki, C., Boffin, H.M.J., Cuypers, J., Stanishev, V., 
Kraicheva, Z. $\&$ Genkov, V., 2004, in Hilditch, R.W., Hensberge, H. and Pavlovski, K., eds, 
Spectroscopically and Spatially Resolving the Components of Close Binary Stars, ASP Conference Series, in Press
\bibitem[Patterson (1998)]{Patterson1998}Patterson, J., 1998, PASP 110, 1132
\bibitem[Patterson (1999)]{Patterson1999}Patterson, J., 1999, in Mineshige S., Wheeler C., eds, 
Disk Instabilities in Close Binary Systems. Universal 
Academic Press, Tokyo, p.61
\bibitem[Patterson (2001)]{Patterson2001}Patterson, J., 2001, PASP 113, 736
\bibitem[Patterson $\&$ Richman (1991)]{PattersonandRichman1991}Patterson, J. $\&$ Richman, H., 1991, 
PASP 103, 735
\bibitem[Patterson et al. (1993)]{Pattersonetal1993}Patterson, J., Thomas, G., Skillman, D.R. $\&$ Diaz, M., 
1993, ApJS 86, 235
\bibitem[Patterson $\&$ Skillman (1994)]{PattersonandSkillman1994}Patterson, J. $\&$ Skillman, D.R., 1994, 
PASP 106, 1141
\bibitem[Patterson et al. (1997)]{Pattersonetal1997}Patterson, J., Kemp, J., Saad, J., Skillman, D.R., 
Harvey, D., Fried, R., Thorstensen, J.R. $\&$ Ashley, R., 1997, PASP 109, 468
\bibitem[Patterson et al. (2003)]{Pattersonetal2003}Patterson, J., Thorstensen, J.R., Kemp, J., Skillman, D.R., 
Vanmunster, T., Harvey, D.A., et al., 2003, PASP 115, 1308
\bibitem[Retter, Leibowitz $\&$ Ofek (1997)]{Retteretal1997}Retter, A., Leibowitz, E.M., $\&$ Ofek, E.O. 1997, 
MNRAS, 286, 745
\bibitem[Retter, Naylor $\&$ Leibowitz (1999)]{Retteretal1999}Retter, A., Naylor, T. $\&$ Leibowitz, E.M., 1999, 
"Novae Crossing the Thermal Stability Line", in {\it Disk Instabilities in Close Binary Systems, 25 Years of 
Disk-Instability Model}, ed. S. Mineshige, $\&$ C. Wheeler (Universal Academy Press Inc.: Tokyo), 91
\bibitem[Retter $\&$ Naylor (2000)]{RetterandNaylor2000}Retter, A. $\&$ Naylor, T. 2000, MNRAS, 319, 510
\bibitem[Retter et al. (2002)]{Retteretal2002}Retter, A., Chou, Y., Bedding, T.R. $\&$ Naylor, T., 2002, 
MNRAS 330, L37
\bibitem[Retter et al. (2003)]{Retteretal2003} Retter, A., Hellier, C., Augusteijn, T., Naylor, T., 
Bedding, T. R., Bembrick, C., McCormick, J., $\&$ Velthuis, F., 2003, MNRAS, 340, 679
\bibitem[Retter, Richards $\&$ Wu (2005)]{Retteretal2005}Retter, A., Richards, M.T. $\&$ Wu, K., 2005, ApJ 621, 417
\bibitem[Ringwald, Thorstensen $\&$ Hamwey (1994)]{Ringwaldetal1994}Ringwald, F.A., Thorstensen, J.R. $\&$ 
Hamwey, R.M., 1994, MNRAS 271, 323
\bibitem[Roberts, Lehar $\&$ Dreher (1987)]{Roberstsetal1987}Roberts, D.H., Lehar, J. $\&$ Dreher, J.W., 1987, AJ 93, 968
\bibitem[Scargle (1982)]{Scargle1982}Scargle, J., 1982, ApJ 263, 835  
\bibitem[Sperl (1998)]{Sperl1998}Sperl, M., 1998, Comm. Astr. Seis., 111
\bibitem[Stanishev et al. (2002)]{Stanishevetal2002}Stanishev, V., Kraicheva, Z., Boffin, H.M.J. $\&$ Genkov, V., 2002, 
A$\&$A, 394, 625
\bibitem[Warner (1995)]{Warner1995}Warner, B. 1995, {\it Cataclysmic Variable Stars}, Cambridge Univ. Press, Cambridge
\bibitem[Warner and Nather (1988)]{WarnerandNather1988}Warner, B. $\&$ Nather, R.E., 1988, IBVS No.3140
\bibitem[Wood, Montgomery $\&$ Simpson (2000)]{Woodetal2000}Wood, M.A., Montgomery, M.M. $\&$ Simpson, J.C., 2000, 
ApJ 535, L39  
\end{thebibliography}
\end{document}